\documentclass[prd,superscriptaddress,nofootinbib,reprint,twocolumn]{revtex4-1}
\usepackage{amsmath,amssymb,bm,epsfig,color,graphicx}
\usepackage[mathscr]{eucal}
\usepackage{slashed}
\usepackage{cancel}
\usepackage[colorlinks=true,linkcolor=blue,citecolor=blue,urlcolor=blue]{hyperref}
\usepackage[caption=false]{subfig}
\usepackage{hyperref}
\usepackage{tikz,xcolor}
\usepackage{amsmath}
\usepackage{slashed}
\usepackage{comment}
\usepackage{cancel}
\usepackage{multirow}

\usepackage{slashed}

\renewcommand{\thefootnote}{\fnsymbol{footnote}}

\newcommand{\bea}{\begin{array}}
\newcommand{\eea}{\end{array}}
\newcommand{\beq}{\begin{eqnarray}}
\newcommand{\eeq}{\end{eqnarray}}

\definecolor{orange}{RGB}{255,100,0}
\definecolor{rosepink}{RGB}{248,100,100}

\begin{document}
%\rightline{preprint number}

\vspace{-0.5cm}

\title{
\vspace{0.5cm}
Inflationary QCD phase diagram
}

%\author{
%Toshifumi Noumi\footnote{
%E-mail address: nomi}\\*[10pt]
%{\it \normalsize
%Graduate School of Arts and Sciences, University of Tokyo, Komaba,\\ Meguro-ku, Tokyo 153-8902,
%Japan} \\*[3pt]
%}

\author{Kohei Fujikura}
\email{kohei.fujikura@yukawa.kyoto-u.ac.jp }
\affiliation{Yukawa Institute for Theoretical Physics, Kyoto University, Kyoto 606-8502, Japan}

\author{Toshifumi Noumi}
\email{tnoumi@g.ecc.u-tokyo.ac.jp}
\affiliation{Graduate School of Arts and Sciences, University of Tokyo, Komaba, Meguro-ku, Tokyo 153-8902,Japan}

\begin{abstract}
    Motivated by the cosmological collider program, which aims to probe high-energy physics through inflation, we investigate the phase diagram of multi-flavor QCD in de Sitter spacetime with a flavor-universal axial chemical potential induced by a rolling inflaton coupled to fermions. We determine the first-order critical line and a critical point as functions of the Hubble parameter and the axial chemical potential, employing an effective description of chiral symmetry breaking within the framework of the Nambu--Jona-Lasinio model. We find that a first-order chiral phase transition may occur during inflation or at its end when the axial chemical potential is sufficiently large and crosses the critical line. This provides a cosmological collider analogue of the QCD phase diagram explored in heavy-ion colliders.
    %\TN{I added some motivations in the abstract. Please check if it is OK?}
    %We investigate \TNmod{the phase diagram of multi-flavor QCD} with a flavor universal axial chemical potential induced by a rolling inflaton coupled to fermions in the de Sitter spacetime. We clarify the (first-order) critical line and a critical point as a function of the Hubble parameter and the axial chemical potential, by employing an effective description of chiral symmetry breaking within the framework of the Nambu–Jona-Lasinio model. A first-order chiral phase transition may take place during or at the end of the inflation when the axial chemical potential is large enough and crosses the critical line.
% Some phenomenological implications are also discussed.
\end{abstract}

\maketitle

\renewcommand{\thefootnote}{\arabic{footnote}}
\setcounter{footnote}{0}

%%%%%%%%%%%%%%%%%%%%%%%%%%%%%%%%%%%%%%%%%%%%%%%%%%%
%\textit{{Introduction.---}} 
%%%%%%%%%%%%%%%%%%%%%%%%%%%%%%%%%%%%%%%%%%%%%%%%%%%
\section{Introduction}

The early universe inflation~\cite{Starobinsky:1980te,Guth:1980zm,Sato:1981qmu} offers a unique and exciting avenue to high energy physics. Inflation as a paradigm of primordial cosmology is strongly supported by cosmic microwave background observations~\cite{2011ApJS..192...18K,Planck:2018vyg} and now we are at the stage to identify the nature of the inflaton and, in turn, utilize inflation as a probe of new physics beyond the Standard Model. Indeed, within the cosmological collider program~\cite{Chen:2009zp, Baumann:2011nk, Noumi:2012vr, Arkani-Hamed:2015bza}, it has been studied how to explore new particles predicted by high energy theories using inflationary observables such as primordial non-Gaussianities~\cite{Maldacena:2002vr,Acquaviva:2002ud}. Given that the Hubble scale $H$ during inflation could be as high as $10^{13}$ GeV, it is important to explore what we can learn about high-energy theories from cosmological observations.

Based on this background, let us take the analogy between inflation and collider experiments one more step further. While detection of new particles is a central role of collider experiments, heavy-ion colliders such as LHC and RHIC have another important role to unveil phase structures of quantum chromo dynamics (QCD) at finite temperature and finite density. This motivates us to explore a possibility to probe phase structures of the particle physics model at the inflation scale in a similar fashion. In fact, the inflationary universe can be regarded as a thermal bath with the Hawking temperature $ T_{\rm dS}=H/2\pi$~\cite{Gibbons:1977mu}, hence it readily provides such an arena. 

Another remarkable fact is that the inflaton velocity $\dot{\phi}$ may play a role of chemical potential~\cite{Cohen:1987vi,Cohen:1988kt}. For example, consider an inflationary scenario where the inflaton $\phi$ is a $CP$-odd scalar that naturally accommodates an axial derivative coupling with a Dirac fermion $\psi$,
\begin{align}
    \mathcal{L}_{\rm int}= \frac{\partial_\mu\phi}{f}\bar{\psi}\gamma^\mu\gamma_5\psi\,,\label{eq:axion derivative coupling}
\end{align}
where $f$ is a decay constant. Since inflaton has a nonzero velocity $\dot{\phi}$ during inflation, this coupling leads to an inflaton-induced {\it effective axial chemical potential},
\begin{align}
\mu_5=\frac{\dot{\phi}}{f}\,,
\end{align}
for the fermion field $\psi$. See Refs.~\cite{Adshead:2015kza,Adshead:2018oaa,Chen:2018xck,Domcke:2018eki,Hook:2019zxa,Wang:2019gbi,Tong:2023krn} for cosmological implications of the inflaton-induced axial chemical potential such as particle production and phase transitions.

In Ref.~\cite{Tong:2023krn}, the phase diagram of a Bardeen-Cooper-Schrieffer (BCS)-like model during inflation was studied identifying the Hubble parameter $H$ and the inflaton velocity $\dot{\phi}$ as a temperature and a chemical potential, respectively. Interestingly, their model accommodates a first-order critical line. Moreover, $H$ and $\dot{\phi}$ time-evolve slowly, so that a first-order phase
transition takes place if the critical line is crossed during inflation, which may leave interesting gravitational wave signals and primordial non-Gaussianities~\cite{Tong:2023krn}. This motivates a comprehensive study of inflationary phase diagrams of particle physics models to explore a role of inflation as a probe of phase structures of high-energy theories.
Since details of the inflaton coupling depend on inflationary scenarios, such studies will, in turn, shed light on the nature of the inflaton as well.
As a benchmark model to initiate this direction, we consider multi-flavor QCD with an inflaton-induced axial chemical potential in the present paper.

The role of axial chemical potential in QCD has been studied in the literature on flat spacetime.
At high temperature, QCD admits the topologically changing process called a strong sphaleron~\cite{McLerran:1990de}, which induces an effective axial chemical potential.
Having in mind application to high-temperature fermion matters with a chiral imbalance, e.g., produced at RHIC, the phase structure of QCD with an axial chemical potential has been extensively investigated on flat spacetime, based on Nambu-Jona-Lasinio model~\cite{Yu:2015hym,Farias:2016let,Lu:2016uwy,Khunjua:2018jmn,Yang:2019lyn},~Polyakov-Nambu-Jona-Lasionio method~\cite{Fukushima:2010fe,Gatto:2011wc,Cui:2016zqp,Azeredo:2024sqc}, linear sigma model coupled to fermions and to the Polyakov loop~\cite{Chernodub:2011fr}, Schwinger-Dyson equation~\cite{Xu:2015vna,Wang:2015tia} and the first-principle lattice simulation~\cite{Braguta:2015zta,Braguta:2015owi}.
For our purpose, it is desirable to generalize these studies to the inflationary QCD phase diagram. In this paper, as a first attempt, we employ the Nambu-Jona-Lasinio (NJL) model~\cite{Nambu:1961tp,Nambu:1961fr} as an effective description of chiral symmetry breaking in multi-flavor QCD on inflationary spacetime (see ref.~\cite{Hatsuda:1994pi} for a review of the NJL model and its application to the ordinary QCD).\footnote{In the absence of the axial chemical potential, the NJL model on the curved space background has been extensively investigated in many literature (See e.g. Ref.~\cite{Inagaki:1997kz} for a review).}

The paper is organized as follows.
We discuss the NJL model in de Sitter space and compute the effective potential of chiral condensates with a mean-field approximation in Sec~\ref{sec:NJL model}.
The inflationary QCD phase diagram is revealed in Sec.~\ref{sec:inflationary QCD phase diagram}.
Section \ref{sec:conclusion and discussion} is devoted to the conclusion and discussion.

\section{Nambu-Jona-Lasinio model in de Sitter space}\label{sec:NJL model}

In this section, we explain a setup of our model, and give an integral representation of the potential of chiral condensate using the Nambu-Jona-Lasinio model with a mean-field approximation.

Our main focus is on the symmetry breaking pattern $SU(N_f)_L\times SU(N_F)_R\times U(1)_V \times U(1)_A\to SU(N_f)_V\times U(1)_V$ with $N_f>1$ being the flavor number of fermions.
We work in the de Sitter spacetime with a Poincar\'e patch whose metric is defined by
\begin{align}
    \mathrm{d}s^2=\frac{-\mathrm{d}\eta^2+\mathrm{d}{\bm x}^2}{\eta^2H^2} \quad (\eta<0).
\end{align}

Let us introduce Dirac fermions $\psi_{c,f}$ with the color index $c=1,2,\cdots,N_c$ and the flavor index $f=1,2,\cdots,N_f$.
In what follows, we assume that fermions are minimally coupled to gravity, and there is a flavor universal fermion coupling with inflaton which is of the form Eq~\eqref{eq:axion derivative coupling}.
We consider a four-Fermi operator, 
\begin{align}
    \mathcal{L}_{\rm NJL}= \frac{G_S}{2} \sum_A\left((\bar{\psi}_{c,f}\lambda^A_{ff'}\psi_{c,f})^2+(\bar{\psi}_{c,f}\lambda^A_{ff'}i\gamma_5\psi_{c,f'})^2\right).\label{eq:fermi interaction}
\end{align}
Here, $\lambda^A$ ($A=0,1,\cdots,N_f-1$) are the generators of $SU(N_f)$, $\lambda^{A=a}$ ($a=1,2,\cdots,N_f-1$) with ${\rm tr}\lambda^A\lambda^B=2\delta^{AB}$ and identity matrix $\lambda^{A=0}= {\bm 1}_{N_f\times N_f}\sqrt{\frac{2}{N_f}}$.
The above operator respects all flavor symmetries.
In our setup, we introduce the following Kobayashi-Maskawa-'t Hooft (KMT) interaction~\cite{Kobayashi:1970ji,Kobayashi:1971qz,tHooft:1976snw,tHooft:1986ooh,Shifman:1978bx},
\begin{align}
    \mathcal{L}_{\rm KMT}=G_{\rm KMT}\left\{\det\left[\bar{\psi}_c(1-\gamma_5)\psi_c\right]+{\rm c.c.}\right\},\label{eq:KMT term}
\end{align}
where the determinant is taken over flavor indices.
In the ordinary QCD, the above interaction is generated by integrating out a topologically number changing process such as the instanotn~\cite{tHooft:1976rip,Gross:1980br}.
This effective operator breaks the axial symmetry $U(1)_A$ down to its discrete subgroup, ${\bf Z}_{2N_f}$.

We also include flavor dependent fermion masses, 
\begin{align}
    \mathcal{L}_{\rm mass}=-\bar{\psi}_{c,f}\mathcal{M}_{ff'}\psi_{c,f'}.\label{eq:fermion masses}
\end{align}
Without loss of generality, one can work in the mass diagonal basis, $\mathcal{M}_{ff'}={\rm diag}(m_1,m_2,\cdots,m_{N_f})$.
The above mass terms explicitly break the $U(1)_A,~SU(N_f)_A$ and $SU(N_f)_V$ symmetries.
When some fermion masses degenerate, a subgroup of $SU(N_f)_V$ flavor symmetry remains.

We investigate the phase structure of this theory using the conventional auxiliary field method~\cite{Reinhardt:1988xu,Ebert:1985kz}.
Treating the inflaton field $\phi$ as the background field with a constant velocity, we consider the partition function with the Lagrangian densities given by \eqref{eq:axion derivative coupling},~\eqref{eq:fermi interaction},~\eqref{eq:KMT term}, and \eqref{eq:fermion masses} and a canonically normalized fermion kinetic term $\mathcal{L}_{\rm kin}$.
Introducing an $N_f\times N_f$ matrix-valued auxiliary field $\Sigma_{ff'}(x)$, the fermion path-integral can be rewritten as the following form,
\begin{align}
    Z=\int D\bar{\psi}D\psi D\Sigma^\dag D\Sigma \,e^{iS_{\rm NJL}},\label{eq:path-integral}
\end{align}
where $S_{\rm NJL}$ is given by 
\begin{align}
    &S_{\rm NJL}=\int\mathrm{d}^4 x\sqrt{-g}\left(\mathcal{L}_{\rm kin}+\mathcal{L}_{\rm mass}+\mathcal{L}_{\rm \Sigma}+\mathcal{L}_{\rm KMT}\right).
\end{align}
In this expression, $\mathcal{L}_{\rm NJL}$ is given by
\begin{align}
    -\mathcal{L}_{\Sigma}=\frac{1}{4G_S}{\rm tr}\left(\Sigma \Sigma^\dag\right)+\bar{\psi}_{c}(P_L\Sigma+P_R\Sigma^\dag)\psi_{c},
\end{align}
where $P_{L(R)}\equiv (1\pm \gamma_5)/2$ is the chiral projection operator to the left (right)-handed fermion.
The kinetic term of fermion fields is canonically normalized, and the external sources of fermions are omitted in the above expressions.
The Bunch-Davies boundary condition is taken for the definition of the path integral.\footnote{A precise definition of the contour of path-integral is unimportant in the following discussion because we only need to compute the fermion propagator
connecting the two identical spacetime points.
%at two same spacetime points.
}

In what follows, we perform a mean-field analysis, $\Sigma(x)\simeq \Sigma$.
A variation with respect to $\Sigma^\dag$ gives
\begin{align}
    ~\Sigma_{ff'} = -4G_S\langle \bar{\psi}_{c,f}P_R\psi_{c,f'}\rangle.\label{eq:saddle-point}
\end{align}
We evaluate the fermion path-integral defined by eq.~\eqref{eq:path-integral} around the saddle-point eq.~\eqref{eq:saddle-point} to obtain the effective potential $V_{\rm eff}[\Sigma]\equiv -i\log Z[\Sigma]/ \int \mathrm{d}^4x\sqrt{-g}$ where we divide the spacetime volume.
Generically, $\Sigma$ contains off-diagonal components because $SU(N_f)_V$ symmetry is explicitly broken by the fermion mass term eq.~\eqref{eq:fermion masses}.
We assume that this explicit breaking is small enough to safely approximate $\Sigma_{ff'}\simeq \Sigma_f\delta_{ff'}$ with $\Sigma_f$ being real.
This approximation can be justified in the large $N_c$ limit~\cite{Ebert:1985kz}.
The KMT operator eq.~\eqref{eq:KMT term} is non-linear with respect to the fermion billinear, and Gaussian fluctuations around eq.~\eqref{eq:saddle-point} are taken into account.
Then the derivative of the effective potential can be formally expressed as
\begin{equation}
\begin{split}
    \frac{\partial V_{\rm eff}}{\partial \Sigma_f}= \frac{\Sigma_f}{2G_S} &+ \frac{\partial M_{f'}}{\partial \Sigma_f}\langle \bar{\psi}_{c,f'}\psi_{c,f'}\rangle \\
    &+ 2(N_F-1)\left(-\frac{1}{2G_S}\right)^{N_F}G_{\rm KMT}\prod_{f''\neq f}\Sigma_{f''}.\label{eq:derivative of effective potential}
\end{split}
\end{equation}
Here, the constituent mass $M_f$ should be evaluated at the saddle-point eq.~\eqref{eq:saddle-point} which is given by 
\begin{align}
    M_f= m_f + \Sigma_f -\left(-\frac{1}{2G_S}\right)^{N_f-1}2G_{\rm KMT}\prod_{f'\neq f}\Sigma_{f'}.
\end{align}
The flavor-diagonal components of the chiral condensate $\langle \bar{\psi}_{c,f}\psi_{c,f}\rangle$ is evaluated by the canonical quantization on the de Sitter space background with an effective axial chemical potential (See also Appendix \ref{appendixA}).
At one-loop order, it is explicitly expressed by the following form.
\begin{align}
\langle \bar{\psi}_{c,f}\psi_{c,f}\rangle
&= -N_c \frac{\tilde{M}_f}{4\pi^2} H^3
   \Bigl[
      F_+(\tilde{\Lambda},\tilde{\Delta}_f,\tilde{\mu}_5)
      + F_-(\tilde{\Lambda},\tilde{\Delta}_f,\tilde{\mu}_5)
\nonumber\\
&\qquad\qquad\qquad
      + \text{c.c.}
   \Bigr].
\label{eq:fermion condensate}
\end{align}
In the above expression, $F_{\pm}$ is defined by
\begin{equation}
\begin{split}
&F_\pm (\tilde{\Lambda}_f,\tilde{\Delta},\tilde{\mu}_5)= \\
&\mp ie^{\mp \pi\tilde{\mu}_5}\int_0^{\tilde{\Lambda}} \mathrm{d}z zW_{\mp 1/2\pm i\tilde{\mu}_5, i\tilde{\Delta}_{f}}(-2iz)W_{\pm 1/2\mp i\tilde{\mu}_5,-i\tilde{\Delta}_{f}}(2iz),\label{eq:F function}
\end{split}
\end{equation}
where all dimensionful parameters are normalized by the Hubble parameter as $\tilde{\mu}_5\equiv \mu_5/H,~\tilde{\Delta}\equiv \sqrt{\tilde{\mu}^2_5+\tilde{M}^2_f}$ and $\tilde{\Lambda}=\Lambda/H$.
We regularized the spatial momentum integration by introducing the three-dimensional sharp cut-off.
$W_{\kappa,\lambda}(z)$ is the Whittaker function.
When minima of the effective potential are non-zero values of $\Sigma_f$, fermions obtain masses by spontaneous breaking of the chiral symmetry.

%%%%%%%%%%%%%%%%%%%%%%%%%%%%%%%%%%%%%%%%%%%%%%%%%%%
%\textit{{Inflationary QCD Phase Diagram.---}} 
%%%%%%%%%%%%%%%%%%%%%%%%%%%%%%%%%%%%%%%%%%%%%%%%%%%
\section{Inflationary QCD Phase Diagram}\label{sec:inflationary QCD phase diagram}

We reveal the QCD phase diagram in the inflationary Universe by solving the gap equation eq.~\eqref{eq:derivative of effective potential}.
This equation is non-linear with respect to the chiral condensate and contains the Whittaker function with imaginary parameters and imaginary argument, which makes it challenging to find the potential minima analytically.
Therefore, our analysis mainly relies on the numerical analysis.
In the numerical computation, we set all physical quantities normalized by the spatial momentum cutoff, $\Lambda$.

For illustration, we choose ratios $m_f/\Lambda$,~ $G_s\Lambda^2$ and $G_{\rm KMT}\Lambda^5$ to be the same as those in the real-world QCD, but an extension to a generic case is straightforward.
More precisely, we consider $N_c=3$ and $N_f=2+1$, where there are two light fermions with degenerate masses, and the relatively heavy fermion.
As we shall see later, the location of the first-order critical line is sensitive to the mass $m_l$ of light fermions.
Therefore, we analyze the cases with a nonzero light fermion mass $m_{l}/\Lambda = 8.7\times 10^{-3}$ and a vanishing light fermion mass $m_l=0$ for comparison.
The heavy fermion mass is always taken to be $m_s/\Lambda = 0.22$.
In either case, it is enough to treat two chiral condensates, $\Sigma={\rm diag}(\Sigma_{l},\Sigma_{l},\Sigma_s)$ thanks to the $SU(2)_V$ flavor symmetry.

NJL model should reproduce physical observables such as meson decay constants and their masses in the flat spacetime limit.
The matching condition leads $G_S=3.67/\Lambda^2,~G_{\rm KMT}=-9.29/\Lambda^5$ in the real world QCD~\cite{Hatsuda:1994pi}.
We assume that these physical parameters do not depend on $H$ and $\mu_5$ other than the KMT term in de Sitter spacetime.
In other word, effects of the spacetime curvature and the effective axial chemical potential are included in the fermion propagator.
For the KMT interaction, we consider two extreme scenarios; (i) $G_{\rm KMT}$ is same as that of the flat-spacetime limit, and (ii) $G_{\rm KMT}=0$, where the topological changing process is significantly suppressed by the spacetime curvature effect.
Although we do not concretely evaluate the instanton solution in the inflationary Universe, we expect that an actual strength of $G_{\rm KMT}$ takes a value between them because the presence of a cosmological horizon may suppress the contribution from instanton whose radius is longer than $H^{-1}$.

We numerically evaluate the derivative of the effective potential for chiral condensates, $\Sigma_l$ and $\Sigma_s$ given by \eqref{eq:derivative of effective potential}.
For a given parameter set, it is not easy to evaluate the effective potential on the $(\Sigma_l,\Sigma_s)$-plane because one needs to perform double integrals for spatial momentum and condensates.
Therefore, we implement the line-search method to search the local minimum.
We give explanations on the numerical implementation in the appendix \ref{appendixB}.
Using this algorithm, we confirm that the position of the minimum of the effective potential in the limit of $H\to 0$ and $\mu_5\to 0$ is in agreement with that of the zero temperature in ref.~\cite{Fukushima:2008wg}.
In our analysis, we identify a first-order critical line as the line where a coexisting phase is found.
When order parameter is continuous, but its first derivative is discontinuous, we interpret it as the point of the second-order phase transition.

%%%%%%%%%%%%%%%%%%%%%%%%%%%%%%%%%%%
\begin{figure}[t!]
    \centering
    \includegraphics[width=8.5cm]{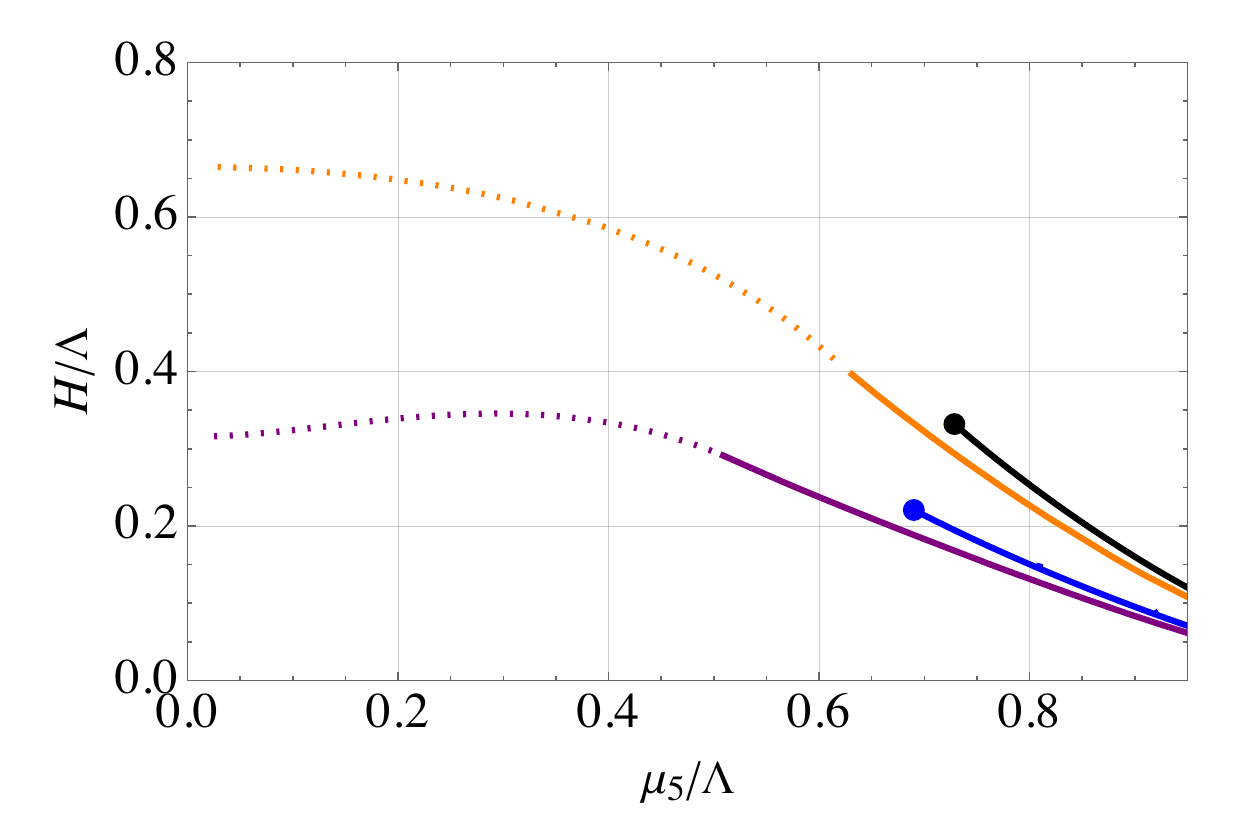}
\caption{Inflationary QCD phase diagram is shown on $(\mu_5/\Lambda,H/\Lambda)$-plane, where $H,~\mu_5$ and $\Lambda$ are the Hubble parameter, the inflaton induced axial chemical potential and three dimensional sharp cutoff, respectively. The black solid curve and the blue dashed curve are first-order critical lines, while large points represent critical points for each cases with non-vanishing light fermion masses.
The solid orange curve (the case (i)) and purple curve (the case (ii)) represent the first-order critical line, and dotted curves correspond to the second-order critical curves for vanishing light fermion masses.
}\label{fig:inflationary QCD phase diagram}
\end{figure}
%%%%%%%%%%%%%%%%%%%%%%%%%%%%%%%%%%%
Fig. \ref{fig:inflationary QCD phase diagram} shows the first-order critical line and its critical point on $(\mu_5/\Lambda,H/\Lambda)$-plane for $m_l/\Lambda=8.7\times 10^{-3}$ with and without the KMT term.
One can see from the figure that chiral condensates are smooth functions with respect to the change of the Hubble parameter in the absence of an axial chemical potential.
Hence, a smooth crossover rather than a definite phase transition is observed in the absence of the inflaton coupling eq.~\eqref{eq:axion derivative coupling}.
In contrast to this result, there can be a first-order phase transition in the presence of the axial potential when $\mu_5/\Lambda \gtrsim 0.7$.
The location of the critical point is less sensitive to the KMT term as can be seen from the figure.
This implies that the first-order phase transition is induced by the axial chemical potential rather than the KMT term in the present setup.

%%%%%%%%%%%%%%%%%%%%%%%%%%%%%%%%%%%
\begin{figure*}[htbp]
    \centering
    \begin{minipage}[t]{0.48\textwidth}
        \centering
        \includegraphics[width=\textwidth]{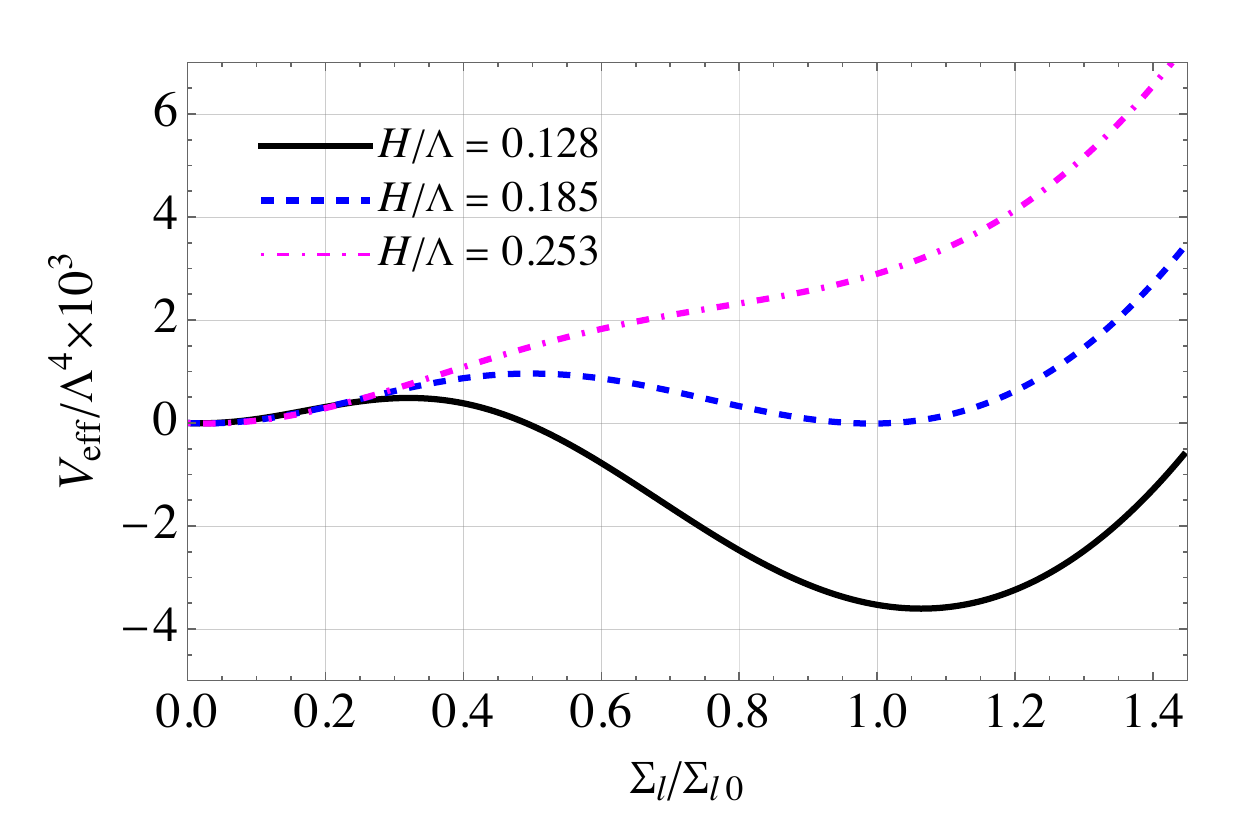}
    \end{minipage}
    \hspace{0.01\textwidth} 
    \begin{minipage}[t]{0.48\textwidth}
        \centering
        \includegraphics[width=\textwidth]{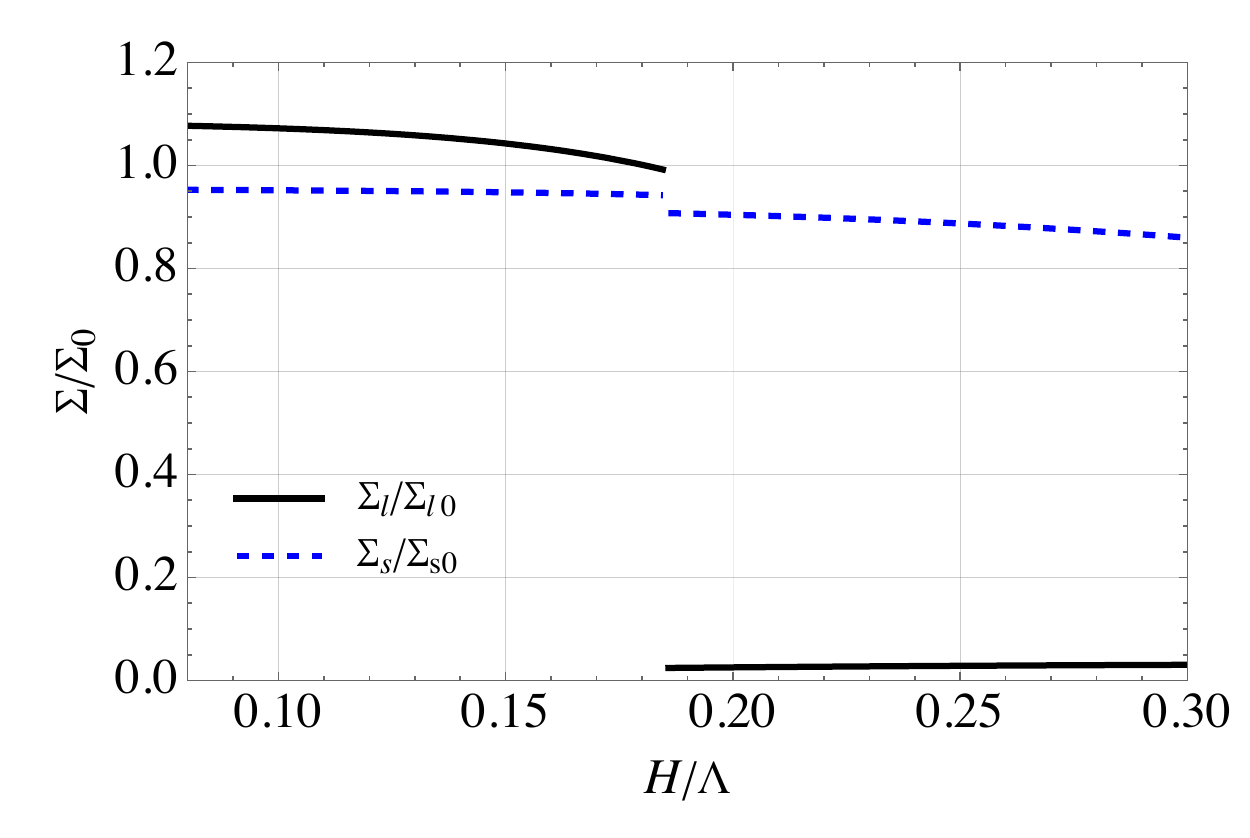}
    \end{minipage}
\caption{Dependence of the effective potential with a fixed $\Sigma_s/\Sigma_{s0}=0.913$ (left) and chiral condensates on the Hubble parameter (right) are shown.
In both panels, an axial chemical potential is taken for $\mu_5/\Lambda=0.871$.
Chiral condensates $\Sigma_l$ and $\Sigma_s$ are normalized by values at $H/\Lambda=0.185$, where two degenerate potential minima exist.\label{fig:around critical line} %their corresponding values evaluated in the flat spacetime limit without the axial chemical potential. \TN{Is $\Sigma_{l0}$ the condensation on flat space with $T=0$ and $\mu=0$?}}\label{fig:potential shape}\label{fig:around critical line}
}
\end{figure*}
%%%%%%%%%%%%%%%%%%%%%%%%%%%%%%%%%%%
We can now take a closer look at the region around the critical line.
Figure \ref{fig:around critical line} shows dependence of effective potential and order parameters on the Hubble parameter around the critical line for a fixed chemical potential $\mu_5/\Lambda =0.871$ in the case (i).
It is clear from the figure that two phases coexist at $H/\Lambda =0.185$.
A discontinuous change of the chiral condensate of light fermions are significant, while the one of the heavy fermion is not so strong because of the relatively large effect from the explicit breaking induced by the mass term, $m_s\gg m_l$.
One can also see from these figures that chiral condensates are smaller, and chiral symmetries almost restore when we make the Hubble parameter larger.
This behavior is also reported in ref.~\cite{Tong:2023krn}, and suggests that the Hubble parameter is identified as the ambient temperature in the flat spacetime.
In fact, the shape of the critical line is qualitatively in agreement with that of the flat spacetime finite-temperature field theory with an identification, $T= H/(2\pi)$, where $T$ is the ambient temperature.
This similarity comes from the fact that the Bunch-Davies boundary condition, which is obtained by an analytic continuation from the Euclidean sphere, is taken for the computation of the fermion condensate.

We now discuss the application to the case with generic QCD whose chiral symmetry breaking pattern is $SU(N_f)_L\times SU(N_f)_R\times U(1)_V\times U(1)_A\to SU(N_f)_V\times U(1)_V$.
Although our benchmark point is motivated by the real world QCD, our analysis can be straightforwardly extended to the generic QCD with arbitrary $N_c$ and $N_f$ with a simple rescaling of three-dimensional sharp cutoff, and change of strength of couplings.
In reality, precise values of four-Fermi coupling, KMT coupling and fermion masses are highly model dependent.
Since a numerical scan for various parameters in the NJL model is cumbersome, we only report the important behavior of the phase diagram in the following.

We numerically confirm that the location of the critical point is quite sensitive to the mass term of the fermion.
This behavior can be understood by the following simple consideration.
In the massless limit, $SU(N_f)_L\times SU(N_f)_R$ symmetry persists, while $U(1)_A$ is explicitly broken down to $\bm{Z}_{2N_f}$ by the KMT term.
Therefore, there exists the phase transition associated with the spontaneous symmetry breaking of them in a strict sense so that the (second-order or first-order) critical line continues to $\mu_5=0$.
We confirm this statement by numerical scan in the (two-flavor) massless limit $m_{l}=0$. See Fig.~\ref{fig:inflationary QCD phase diagram}.
The location of the critical point moves toward a larger $\mu_5$, and eventually it disappears when we make the fermion mass heavier.
Therefore, it is very important to include a fermion mass term to analyze the phase diagram.

%%%%%%%%%%%%%%%%%%%%%%%%%%%%%%%%%%%%%%%%%%%%%%%%%%%
\section{Conclusion and Discussion}\label{sec:conclusion and discussion}
%%%%%%%%%%%%%%%%%%%%%%%%%%%%%%%%%%%%%%%%%%%%%%%%%%%
We have investigated the inflationary QCD phase diagram utilizing the Nambu-Jona-Lasinio model in the inflationary universe with the effective axial chemical potential induced by the rolling inflaton.
The phase diagram is shown in Fig.~\ref{fig:inflationary QCD phase diagram}.
For sufficiently small fermion masses and a large effective axial chemical potential, there is a first-order critical line on the $(\mu_5,H)$-plane in a mean-field analysis.

Although we used the effective description of the NJL model and a mean-field analysis, it is difficult to justify them in the framework of effective theory.
Furthermore, even if we make these assumptions, our result still depends on the regularization scheme of fermion condensates.
In fact, there are lot of discussions on the regularization scheme dependence of the phase diagram with an axial chemical potential in the flat spacetime~\cite{Yu:2015hym,Farias:2016let,Azeredo:2024sqc}.
Therefore, an analysis based on the first-principle approach such as the lattice simulation may be required.

Some comments on interesting future directions are in order.
Firstly, a large non-Gaussian signature may be induced around the critical point because a soft mode can be excited.
A quantitative estimate requires to reveal the dispersion relation of the auxiliary fields $\Sigma_f$ in the condensed phase on the de Sitter space background.
Secondly, it is interesting to concretely derive the gravitational wave signals from first-order phase transitions by specifying the inflation models leading to the coupling eq.~\eqref{eq:axion derivative coupling} in QCD or QCD-like theories.
Finally, we have neglected the back-reaction effect of particle production triggered by the rolling inflaton.
For instance, there can be a coupling of the axion with the $U(1)$ as well as the non-Abelian gauge fields in the generic setup.
It is known that a helical magnetic field is produced in axion inflation models~\cite{Adshead:2016iae}.
When fermions carry additional gauge quantum numbers, we should certainly include this effect in the analysis in the phase diagram.

%%%%%%%%%%%%%%%%%%%%%%%%%%%%%%%%%%%%%%%%%%%%%%%%%%%
\section*{Acknowledgement} 
%%%%%%%%%%%%%%%%%%%%%%%%%%%%%%%%%%%%%%%%%%%%%%%%%%%
K.F. thanks Takashi Hiramatsu for the valuable discussion on an evaluation of the Whittaker function with imaginary parameters and imaginary argument.
The authors thank Yoshimasa Hidaka for a fruitful discussion.
K.F. is supported
by JSPS Grant-in-Aid for Research Fellows Grant No.22J00345 and by JST CREST Grant No. JPMJCR24I3, Japan.
T.N. is also supported by JSPS KAKENHI Grant No. JP22H01220 and MEXT
KAKENHI Grant No. JP21H05184 and No. JP23H04007.

\bibliographystyle{JHEP}
\bibliography{ref}

\appendix

\section{Quantization of fermion in the inflationary Universe}\label{appendixA}

In this appendix, quantization of a fermion field is studied on a four-dimensional de Sitter space background including the axial chemical potential induced by inflaton.
The metric of the de Sitter space is defined by
\begin{equation}
\begin{split}
	\mathrm{d}s^2= a^2(\eta)(-\mathrm{d}\eta^2+\mathrm{d}{\bm x}^2),~a(\eta) = -\dfrac{1}{H\eta}\quad(\eta <0)\label{eq:metric},
\end{split}
\end{equation}
where $\bm{x}=(x^1,x^2,x^3)$ denotes the three-dimensional spatial vector and $H$ is the Hubble parameter.
$\eta$ is the conformal time.

In the mean-field and Gaussian approximations, the fermion condensate is expressed by
\begin{align}
    \langle\bar{\psi}(\eta,{\bm x})\psi(\eta,{\bm x})\rangle =\int D\bar{\psi}D\psi \,\bar{\psi}(\eta,{\bm x})\psi(\eta,{\bm x}) e^{iS_{F
    }},
\end{align}
\begin{equation}
\begin{aligned}
S_F
= \int \mathrm{d}^4x \,\sqrt{-g}\Bigl[
 &-\bar{\psi} i \gamma^a e_a^{\,\mu}\nabla_\mu \psi
 -\bar{\psi} M\psi \\
&
 +\frac{\partial_\mu \phi}{f} e^\mu_{\ a}\bar{\psi}\gamma^a\gamma_5\psi
\Bigr].
\end{aligned}
\end{equation}
Here, $e_a^\mu$ ($a=0,1,2,3$) is the vierbein defined by
\begin{align}
    &g_{\mu\nu} = e_\mu^ae_\nu^b\eta_{ab},~\eta={\rm diag}(-1,+1,+1,+1),\\
    &e_{\mu a} e^\mu_{b} =\eta_{ab},~e_\mu^a e_{\nu a}=g_{\mu\nu}.
\end{align}
The covariant derivative for a Dirac fermion is defined by~\cite{Collas:2018jfx}
\begin{align}
    \nabla_\mu&=\partial_\mu +\frac{i}{2}(\omega_\mu)_{ab}\Sigma^{ab},\\
	\Sigma^{ab}&=\dfrac{i}{4} [\gamma^a,\gamma^b].
\end{align}
Here, $[A,B]=AB-BA$ is the standard commuter.
Gamma matrices can be expressed by 
\begin{align}
	\gamma^a\equiv
	\begin{pmatrix}
		0&\sigma^a \\
		\bar{\sigma}^a & 0
	\end{pmatrix},~
	\gamma^5=i\gamma^0\gamma^1\gamma^2\gamma^3=
	\begin{pmatrix}
		1&0\\
		0&-1
	\end{pmatrix},\\
    \sigma^a_{\alpha\dot{\alpha}} = (-{\mathbf 1}_{2\times 2},\sigma^i),~\bar{\sigma}^{a\dot{\alpha}\alpha}= (-{\mathbf 1}_{2\times 2},-\sigma^i),
\end{align}
where $\alpha,\dot{\alpha}=1,2$ denote the $SL(2,\mathbf{C})$ indices.

The Bunch-Davies boundary condition is taken in the definition of the above path integral, where only the positive frequency mode with respect to the conformal time contributes in the remote past.
Since the action is quadratic functional for fermion fields, one can precisely perform the path integral.
We explicitly compute the $\langle \bar{\psi}(\eta,{\bm x})\psi(\eta,{\bm x})\rangle$ in the following.

The spin connection $\omega_{ab}=-\omega_{ba}$ is determined by the following relation:
\begin{align}
	\mathrm{d}e^{a}=e_b\,\wedge\omega^{ab}.
\end{align}
In the coordinate system eq.~\eqref{eq:metric}, one obtains
\begin{align}
	(\omega_\mu)_{[0i]} = -\dfrac{a'}{a^2}(e_\mu)_i=-\dfrac{a'}{a}\delta_{\mu i},~(i=1,2,3).
\end{align}
$[ij]$ represents the anti-symmetric components of indices $i$ and $j$.
Other components are zero.
Using the relations,
\begin{align}
	\frac{i}{2}\gamma^a (e_a)^\mu(\omega_\mu)_{cd} \Sigma^{cd}= \frac{3}{2}\dfrac{a'}{a^2}\gamma^0,\\
	\slashed{\partial} = \gamma^a(e_a)^\mu\partial_\mu=\dfrac{1}{a}\gamma^a (\partial_a),
\end{align}
the Dirac operator can be expressed as
\begin{align}
	\gamma^ae_a^\mu \nabla_\mu &=\dfrac{1}{a}\gamma^a(\partial_a)+\frac{3}{2}H\gamma^0.
\end{align}

The inflaton is assumed to be homogeneous and has a non-vanishing velocity, $\partial_\eta \phi\neq 0$.
We introduce the axial chemical potential in the local Lorentz frame defined by $\mu_5=(\partial_{\eta} \phi)e^{\eta}_0/f=(\partial_{\eta} \phi)/(a(\eta)f)$.
The $\eta$-dependence on $\mu_5$ is neglected in the following discussion, assuming that the second slaw-roll approximation on the inflaton sector is valid.

We expand the fermion field as follows.
\begin{align}
    &\psi=a^{-3/2}\begin{pmatrix}
    \chi_\alpha\\
    \xi^{\dot{\alpha}}
        \end{pmatrix},\nonumber\\
    &\chi_\alpha= \int \dfrac{\mathrm{d}^3{\bm k}}{(2\pi)^3}\left[h_{\alpha s}a^s_{\bm k}u_s(k,\eta)e^{i{\bm k}\cdot{\bm x}}+h_{\alpha s}(b^\dag)^s_{\bm k}v_s^*(k,\eta)e^{-i{\bm k}\cdot{\bm x}}\right],\\
    &\xi^{\dot{\alpha}}=\int \dfrac{\mathrm{d}^3{\bm k}}{(2\pi)^3}\left[(h^*)_{\alpha s}a^s_{\bm k}v_s(k,\eta)e^{i{\bm k}\cdot{\bm x}}+(h^*)_{\alpha s}(b^\dag)^s_{\bm k}u_s^*(k,\eta)e^{-i{\bm k}\cdot{\bm x}}\right]\nonumber
\end{align}
In these expressions, $\chi_\alpha$ and $\xi^{\dot{\alpha}}$ are left and right-handed Weyl spinors, respectively.
The index $s$ is the two-component helicity, and $h_{\alpha s}$ is the $2\times 2$ unitary matrix satisfying following condition,
\begin{align}
    ({\bm \sigma}\cdot{\bm k})h = s|k|h,~s=\pm 1.
\end{align}
$a^s_{\bm k},~(a^\dag)^s_{\bm k}$ and $b^s_{\bm k},~(b^\dag)^s_{\bm k}$ are creations (annihilations) operators of particle and anti-particle, respectively.
They satisfy standard canonical anti-commutation relations,
\begin{align}
     \{a^{s_1}_{{\bm k}_1},(a^\dag)^{s_2}_{{\bm k}_2}\}=\{b^{s_1}_{{\bm k}_1},(b^\dag)^{s_2}_{{\bm k}_2}\}=(2\pi)^3\delta^{s_1s_2}\delta({\bm k}_1-{\bm k}_2).
\end{align}
Other anti-commutations are zero.
A vacuum condition is taken to be $a^s_{\rm k}|{\rm BD}\rangle=b^s_{\rm k}|{\rm BD}\rangle =0$.

The mode functions $u_s(k,\eta)$ and $v_s(k,\eta)$ are solutions of $-i\mathcal{D}u(k,\eta)=Mv(k,\eta)$ and $-i\bar{\mathcal{D}}v(k,\eta)=Mu(k,\eta)$, where operators $\mathcal{D}$ and $\bar{\mathcal{D}}$ are given by
\begin{align}
		&\mathcal{D}=\frac{1}{a}\sigma^a\partial_a-i\mu_5\sigma^0,\\
		&\bar{\mathcal{D}}=\frac{1}{a}\bar{\sigma}^a\partial_a+i\mu_5\bar{\sigma}^0.
\end{align}
Hence $u_s(k,\eta)$ and $v_s(k,\eta)$ satisfy $\bar{\mathcal{D}}\mathcal{D}u_s=-M^2u_s$ and $\mathcal{D}\bar{\mathcal{D}}v_s=-M^2v_s$, namely,
\begin{align}
    &\frac{1}{a^2}u_s''(k,\eta)
    -\frac{H}{a}u_s'(k,\eta)\nonumber\\
    &+\left(\left[-is\,\frac{H}{a}k
           +\left(\frac{k}{a}+s\mu_5\right)^2
      \right]
    +M^2
  \right)u_s(k,\eta)=0,\\
  &\frac{1}{a^2}v_s''(k,\eta)
    -\frac{H}{a}v_s'(k,\eta)\nonumber\\
    &+\left(\left[is\,\frac{H}{a}k
           +\left(\frac{k}{a}+s\mu_5\right)^2
      \right]
    +M^2
  \right)v_s(k,\eta)=0.
\end{align}
Introducing a dimensionless variable $z=-2k\eta$, these differential equations can be rewritten into the following form,
\begin{align}
    &u_s''(z)+\dfrac{1}{z}u_s'(z)\nonumber\\
    &+\left[-s\dfrac{i}{2z}
    +\left(-\frac{1}{2}-s\dfrac{\tilde{\mu}_5}{z}\right)^2+\dfrac{\tilde{M}^2}{z^2}\right]u_s(z)=0,\\
    &v_s''(z)+\dfrac{1}{z}v_s'(z)\nonumber\\
    &+\left[s\dfrac{i}{2z}
    +\left(-\frac{1}{2}-s\dfrac{\tilde{\mu}_5}{z}\right)^2+\dfrac{\tilde{M}^2}{z^2}\right]v_s(z)=0.
\end{align}
Here, we have introduced dimensionless variables $\tilde{\mu}_5=\mu_5/H$, $\tilde{M}=M/H$ and  $\tilde{\kappa}={\sqrt{\tilde{M}^2+\tilde{\mu}_5^2}}$.
There are two solutions of the above differential equations.
The mode functions behave as the positive (negative) plane waves in the remote past $z\to\infty$ are given by 
\begin{align}
    &u_+= \frac{\tilde{M}}{\sqrt{z}}e^{-\pi\tilde{\mu}_5/2}W_{-\frac{1}{2}+i\tilde{\mu}_5,i\tilde{\kappa}}(-iz),\\
    &u_-=\frac{i}{\sqrt{z}}e^{+\pi\tilde{\mu}_5/2}W_{\frac{1}{2}-i\tilde{\mu}_5,i\tilde{\kappa}}(-iz),\\
    &v_+=\frac{i}{\sqrt{z}}e^{-\pi\tilde{\mu}_5/2}W_{\frac{1}{2}+i\tilde{\mu}_5,i\tilde{\kappa}}(-iz),\\
    &v_-=\frac{\tilde{M}}{\sqrt{z}}e^{\pi\tilde{\mu}_5/2}W_{-\frac{1}{2}-i\tilde{\mu}_5,i\tilde{\kappa}}(-iz).
\end{align}
Overall normalizations are fixed by the canonical anti-commutation relation of a fermion field.
From these expressions, one can compute $\langle{\rm BD}|\bar{\psi}\psi|{\rm BD}\rangle$ under the Bunch-Davies vacuum condition.

\section{Details of numerical computations}\label{appendixB}

In this appendix, we explain a numerical algorithm for searching potential minima.
The potential minima should satisfy the condition $\partial V_{\rm eff}/\partial \Sigma_f =0$, where $\partial V_{\rm eff}/\partial \Sigma_f$ ($f=1,2$) is defined by a set of equations~\eqref{eq:derivative of effective potential},~\eqref{eq:fermion condensate}, and~\eqref{eq:F function}.
A main bottleneck in the numerical computation is to evaluate the Whittaker function $W_{\kappa,\mu}(z)$ with complex parameters $\kappa,\mu\in \mathbf{C}$ and argument $z\in \mathbf{C}$.
The Whittaker function has a brach point at $z=0$ and an essential singular point at $z=\infty$.
In what follows, we only consider the region $|\arg z|<\pi$ and $|z|<\infty$.

The Whitakker function can be expressed by
\begin{align}
    W_{\kappa,\mu}(z)=e^{-z/2}z^{1/2+\mu}U\left(\frac{1}{2}+\mu-\kappa,1+2\mu,z\right),~\label{eq:Whittaker function}
\end{align}
where $U(a_u,b_u;z)$ is the Tricomi (confluent hypergeometric) function.
$U(a_u,b_u;z)$ can be further expressed by a linear combination of the generalized hypergeometric (Kummer's) function ${}_1F_1 (a_F,b_F;z)$ as
\begin{equation}
\begin{split}
    &U(a_u,b_u,z)=\dfrac{\Gamma(1-b_u)}{\Gamma(a_u+b_u-1)}{}_1F_1(a_u,b_u;z)\\
    &+\dfrac{\Gamma(b_u-1)}{\Gamma(a_u)}z^{1-b_u}{}_1F_1(a_u-b_u+1,2-b_u;z),\label{eq:Tricomi function}
\end{split}
\end{equation}
where $\Gamma(a_\Gamma)$ is the Gamma function defiend by 
\begin{align}
    \Gamma(a_\Gamma)= \int \mathrm{d}t\,t^{a_\Gamma-1}e^{-t}.\label{eq:Gamma function}
\end{align}
The ${}_1F_1(a_F,b_F;z)$ admits the following series representation,
\begin{align}
{}_1F_1(a_F,b_F;z)=\sum_{n=0}^{\infty}\dfrac{(a_F)_n z^n}{(b_F)_n n!},~(|\arg z|<\pi,~|z|<\infty).
\end{align}
In the above expression, $(x)_n=x(x+1)\cdots (x+n-1)$ is the Pochhammer symbol.
The Whittaker function is evaluated by the above series expansion rather than a solution of a differential equation.

In a real arithmetic, a certain truncation of the series expansion is required.
We evaluate ${}_1F_1(a_F,b_F;z)$ approximately by
\begin{align}
    {}_1F_1(a_F,b_F;
    z,N_t) = \sum_{n=0}^{N_t}\dfrac{(a_F)_n z^n}{(b_F)_n n!},~(N_t>1).\label{eq:approximate series expansion}
\end{align}
We choose $N_t$ so that the error of the above expression is smaller than the desired accuracy $\epsilon$,
\begin{align}
    |{}_1F_1(a_F,b_F;
    z,N_t+1)-{}_1F_1((a_F,b_F;
    z,N_t)|<\epsilon.
\end{align}
We use $\epsilon=10^{-14}$, which is sufficient for our purpose.
Although the series expansion eq.~\eqref{eq:approximate series expansion} converges for finite $a,b,z\in \mathbb{C}$, it suffers from the round error for $|z|\gg 1$.\footnote{The flat spacetime limit corresponds to $|\mu|,~|z|\to \infty$ where a significant round error arises. In the opposite limit where all mass scales are smaller than the Hubble parameter, convergence of series expansion is good.}
For this, we perform arbitrary precision arithmetic to evaluate ${}_1F_1(a_F,b_F;z,N_t)$ in this region.
A fast computation of $\Gamma(a_\Gamma)$ defined by eq.~\eqref{eq:Gamma function} is also required to evaluate the Whittaker function.
This can be achieved using the standard Lanczos approximation.
The integral over the momentum of eq.~\eqref{eq:F function} is performed by the Gauss-Legendre quadrature.
Then we can compute $\partial V_{\rm eff}/\partial \Sigma_f$ on a standard laptop computer.

However, the calculation of the effective potential $V_{\rm eff}(\Sigma_f)$ is still demanding because integration with respect to $\Sigma_f$ is required.
Therefore we perform the line-search method described below, which only requires local information of the effective potential.
\begin{itemize}
    \item (i) Randomly generate the initial configuration, $\Sigma_{f({\rm old})}$.
    
    \item (ii) Update the field configuration by
    \begin{align}
    \Sigma_{f'({\rm new})} = \Sigma_{f'({\rm old})}-\left.\frac{\partial V_{\rm eff}}{\partial\Sigma_{f'}}\right|_{\Sigma_f=\Sigma_{f({\rm old})}}\left(\left|\frac{\partial^2V_{\rm eff}}{\partial\Sigma_{f'}^2}\right|\right)_{\Sigma_f=\Sigma_{f({\rm old})}}^{-1}.
    \end{align}
    
    \item (iii) Back to the step (ii) with the replacement, $\Sigma_{f({\rm old})}\to \Sigma_{f({\rm new})}$ until $\sum_f|\Sigma_{f({\rm new})}-\Sigma_{f({\rm old})}|<\varepsilon$ is satisfied, where $\varepsilon$ is the desired precision.
\end{itemize}
In our numerical computation $\varepsilon=10^{-3}$ is used.

A manipulation of step (ii) is motivated by the following consideration.
We would like to optimize $V_{\rm eff}(\Sigma_f+\lambda_f)$ where $\lambda_f$ is the variational vector which optimizes $V_{\rm eff}$.
Using $V_{\rm eff}(\Sigma_f+\lambda_f)\simeq V_{\rm eff}(\Sigma_f)+\lambda_{f'}\partial_{\Sigma_{f'}}V_{\rm eff}+\lambda_{f'}^2(\partial^2_{\Sigma_{f'}}V_{\rm eff})/2$ by neglecting the flavor mixing effect and higher-order terms with the assumption $\partial^2_{\Sigma_f}V_{\rm eff}>0$, one can approximately find the optimized vector as the solution of the following problem.
\begin{align}
    \dfrac{\mathrm{d}}{\mathrm{d}\lambda_f}V_{\rm eff}(\Sigma_f+\lambda_f)=0.
\end{align}
The solution is therefore given by $\lambda_f = -\partial_{\Sigma_f}V_{\rm eff}/(\partial^2_{\Sigma_f}V_{\rm eff})$.
In the actual calculation, $\partial_{\Sigma_f}^2V_{\rm eff}<0$ is occasionally realized.
In this case, a numerical instability arises, which can be voided by the replacement of $\partial_{\Sigma_f}^2V_{\rm eff}\to |\partial_{\Sigma_f}^2V_{\rm eff}|$ such that $\lambda_f$ always ensures the decrease direction of the target function $V_{\rm eff}$.
It turns out that this numerical algorithm is very efficient.
In particular, one finds the local minima within 10 iterations in most of the parameter region.

Only the exception is the region around the critical point, where the first-order critical terminates.
This simply because the correlation length $\xi$ (which is roughly given by $\sqrt{(\partial^2_{\Sigma_f}V_{\rm eff})^{-1}}$ under the mean-field approximation) diverges close to the critical point.
Hence, the convergence of the above algorithm becomes worse and worse as we approach the critical point.
Nevertheless, we can successfully find the local minima by using a neighboring solution as the initial guess.

Although the line-search problem is quite powerful in searching for local minima of the target function, one cannot find the global minimum.
To clarify the global minimum, we perform line integration with respect to $\Sigma_f$ along the contour that connects different local minima.
Line integration is performed by the fourth-order Runge-Kutta method.
Figures \ref{fig:inflationary QCD phase diagram} and \ref{fig:around critical line} are generated using the algorithm explained in this appendix.

\end{document}